\documentclass[letterpaper,11pt]{article}

\pdfoutput=1

\usepackage{mathtools}               				   

\usepackage[protrusion=true,expansion=true]{microtype} 
\usepackage[T1]{fontenc}                               
\usepackage{lmodern}

\usepackage{times}									   
\usepackage{txfonts}								   

\usepackage[english]{babel}							   

\usepackage{setspace}                
\usepackage{graphicx}                
\usepackage{float}                   
\usepackage{hyperref}                
\usepackage[margin=1.2in]{geometry}  
\usepackage[short]{datetime}         
\usepackage[labelfont=bf]{caption}   
\usepackage[bottom]{footmisc}        
\usepackage{authblk}                 

\usepackage[compact]{titlesec}
\titleformat*{\section}{\sffamily\normalsize\bfseries\uppercase}
\titlespacing*{\section}{0pt}{1.5ex}{0ex}
\titleformat*{\subsection}{\sffamily\normalsize\bfseries}
\titlespacing*{\subsection}{0pt}{0ex}{0ex}
\titleformat*{\subsubsection}{\sffamily\normalsize\itshape}
\titlespacing*{\subsubsection}{0pt}{0ex}{0ex}

\makeatletter
\renewcommand{\maketitle}{
\begin{flushleft}       
\vspace*{5mm}
\MakeUppercase{\Large\sffamily\bfseries\@title}   
\vspace{15mm}\\         
{\normalsize\sffamily\@author}        
\end{flushleft}
}
\makeatother

\usepackage{lipsum}                        
\usepackage[utf8]{inputenc}
\usepackage[T1]{fontenc}
\usepackage{amssymb}
\usepackage{nicefrac}
\usepackage{mathtools}
\usepackage{cuted}
\usepackage{upgreek}
\usepackage{amsmath}
\usepackage{graphicx}
\usepackage{hyperref}
\usepackage{xcolor}

\title{Mono-crystalline gold platelets: A high quality platform for surface plasmon polaritons}

  \author[1,2,*]{Korbinian J. Kaltenecker}
  \author[3]{Enno Krauss}
	\author[1,2]{Laura Casses}
  \author[1,2]{Mathias Geisler}
  \author[3]{Bert Hecht}
  \author[2,4,5]{N. Asger Mortensen}
  \author[1,2]{Peter Uhd Jepsen}
  \author[1,2]{Nicolas Stenger}

	\affil[1]{Department of Photonics Engineering, Technical University of Denmark, DK-2800 Kongens~Lyngby, Denmark}
  \affil[2]{Center for Nanostructured Graphene, Technical University of Denmark, DK-2800 Kongens~Lyngby, Denmark}
  \affil[3]{Nano-Optics and Biophotonics Group, Experimentelle Physik 5, Physikalisches Institut, Universit\"at W\"urzburg, D-97074 W\"urzburg, Germany}
  \affil[4]{Center for Nano Optics, University of Southern Denmark, DK-5230 Odense M, Denmark}
	\affil[5]{Danish Institute for Advanced Study, University of Southern Denmark, DK-5230 Odense~M, Denmark}
\begin{document}

\begin{singlespace}

\maketitle

\begin{flushleft}


$^{*}$\textbf{Corresponding Author}: Korbinian J. Kaltenecker, DTU Fotonik, email: \url{korkal@dtu.dk}


\today

\end{flushleft}

\end{singlespace}



\section*{Abstract}

We use mono-crystalline gold platelets with ultra-smooth surfaces and superior plasmonic properties to investigate the formation of interference patterns caused by surface plasmon polaritons (SPPs) with scattering-type scanning near-field microscopy (s-SNOM) at 521~nm and 633~nm. By applying a Fourier analysis approach, we can identify and separate several signal channels related to SPPs launched and scattered by the AFM tip and the edges of the platelet. Especially at the excitation wavelength of 633~nm, we can isolate a region in the center of the platelets where we find only contributions of SPPs which are launched by the tip and reflected at the edges. These signatures are used to determine the SPP wavelength of $\lambda_{SPP}=606$ nm in good agreement with theoretical predictions. Furthermore, we were still able to measure SPP signals after 20~$\upmu$m propagation, which demonstrates impressively the superior plasmonic quality of these mono-crystalline gold platelets.
 

\section*{Introduction}

In the last decade the plasmonic community has put tremendous efforts to find new plasmonic materials \cite{naik2013alternative} and to improve the quality of noble metallic films \cite{mcpeak2015plasmonic,wu2015single,krauss2018controlled} to reduce Ohmic losses and surface scattering in order to achieve longer surface plasmon polaritons (SPPs) propagation and stronger plasmonic resonances for better enhanced field confinements. Improved deposition recipes and template stripping were used to significantly improve the surface roughness of noble metallic films such as gold, silver, copper and aluminum \cite{mcpeak2015plasmonic}. Alternatively, chemically grown colloidal gold platelets \cite{wu2015single,hoffmann2016new,krauss2018controlled,Boroviks:18} with atomically smooth surfaces were synthesized and have proven their superiority in terms of stronger plasmonic resonances \cite{Mejard:17,huang2010atomically} and longer propagation in plasmonic circuits \cite{huang2010atomically}. These gold platelets have enabled fascinating new explorations of fundamental plasmonic properties revealed by two-photon photoemission electron microscopy (2P-PEEM) \cite{spektor2017revealing,frank2017short} and improved plasmonic device performances such as for novel light generation applications \cite{kern2015electrically} and quantum light guiding \cite{siampour2019unidirectional}. Other mono-crystalline gold structures such as gold tappers demonstrated improved nanofocusing properties and second harmonic generation under pulsed illumination \cite{schmidt2012adiabatic}.  

In parallel to these developments in material synthesis and fabrications, scattering-type scanning near-field optical microscopy (s-SNOM) \cite{keilmann2004near} has proven to be an unique and versatile tool to explore the propagation of polaritons \cite{low2017polaritons,basov2016polaritons} from the terahertz \cite{eisele2014ultrafast, mastel2017terahertz} to the visible \cite{tsesses2018optical, li2014origin} region of the electromagnetic spectrum with a few tens of nanometer resolution only. Recently, s-SNOM has been instrumental in the excitation and mapping of plasmons in graphene \cite{chen2012optical, fei2012gate} and van der Waals heterostructures \cite{woessner2015highly}, as well as surface phonon-polaritons in polar dielectric low-dimensional materials \cite{dai2014tunable, yoxall2015direct,ma2018plane} at mid-infrared (MIR) frequencies. In the near-infrared and visible range, s-SNOM has been used to measure the amplitude and phase of gap plasmons \cite{andryieuski2014direct} and to prove the existence of SPP interference patterns with exotic properties \cite{tsesses2018optical}.   

In the visible, the observation of SPP at the surface of metallic films with s-SNOM is more challenging. These challenges originate mainly from intrinsic material and propagation properties of SPP on noble metals with, in addition, technical limitations inherent to s-SNOM. From the material properties perspective, the optical properties of gold or silver suffer from intraband (Landau damping) and interband losses \cite{khurgin2015deal}, limiting the propagation length of SPP to a few micrometers only at visible frequencies. Moreover, the confinement of SPP in noble metals is less pronounced than plasmons in graphene in the MIR \cite{chen2012optical, fei2012gate, woessner2015highly}. This difference can explain the excitation of SPPs at the edge of a gold film \cite{li2014origin, walla2018anisotropic} (edge-launched SPPs) and so-called tip-reflected edge-launched SPPs, recently identified at NIR frequencies on gold surfaces \cite{walla2018anisotropic}. Edge-launched surface waves in the visible have also been investigated on MoS$_2$ where different kinds of surface waves (unbound cylindrical, Zenneck and Zenneck-type waves) contribute to the measured signatures \cite{babicheva2018near}. These channels of excitation are usually negligible in graphene due to the large confinement of these excitations and the main excitation channel of plasmons comes from plasmons launched at the apex of the metallic tip, also called tip-launched SPPs. This excitation channel is of special interest because it acts as a SPP point source that can selectively irradiate a structure of interest on a metal film such as inhomogeneities \cite{hecht1996local}. From a technical point of view, focusing the diffraction-limited spot of the excitation laser onto the metallic tip is more difficult at visible frequency than at MIR frequency due to a smaller size of the illumination spot. Moreover, the signal on the detector $S_{total}$ strongly depends on the shape of the metallic tip and the direction of the illumination \cite{walla2018anisotropic}. All these contributions will interfere directly or indirectly in the far-field to contribute to the final signal $S_{total}$. This interference can lead to complex and intricate patterns making the retrieval of the wavelength and propagation length of SPP in the visible challenging in reflection mode.

These interference effects are reminiscent to the interference patterns of edge-launched SPPs observed in 2P-PEEM \cite{dabrowski2017ultrafast}, which have been thoroughly studied in Ag \cite{dai2018ultrafast,buckanie2013interaction} and Au \cite{gong2015ultrafast,frank2017short} nanoplatelets. However, in the case of s-SNOM the tip not only probes the nearfield but also contributes to additional SPP excitation pathways (tip-launched, tip-reflected edge-launched). We are furthermore sensitive to fields which are scattered from different locations (tip and edge) and interfere at the detector. All these additional effects lead to modified interference patterns compared to the ones observed in 2P-PEEM.

In this work we use high quality mono-crystalline gold platelets with ultra-smooth surfaces \cite{wu2015single}, with reduced losses to explore the launching mechanism of SPP by a metallic tip at two wavelengths in the visible. We investigate the interference mechanism leading to complex interference patterns and confirm the anisotropy of the launching process observed by Walla, \textit{et al.} \cite{walla2018anisotropic}. Furthermore, due to the exceptional plasmonic properties of our mono-crystalline gold platelets, we demonstrate how to retrieve the wavelength of the SPPs excited by 633~nm laser beam with high confidence. We also measure clear signals of SPPs after a propagation distance of at least 20~$\upmu$m. For excitations at 521 nm, we observe strong damping due to the onset of interband transitions in the gold platelets. This work demonstrates that the combination of mono-crystalline platelets with s-SNOM can be applied to directly observe and characterize fundamental properties of SPPs in the visible.

\section*{Experimental Details}

\begin{figure}[htb!]
\centering
	\includegraphics[width=0.50\textwidth]{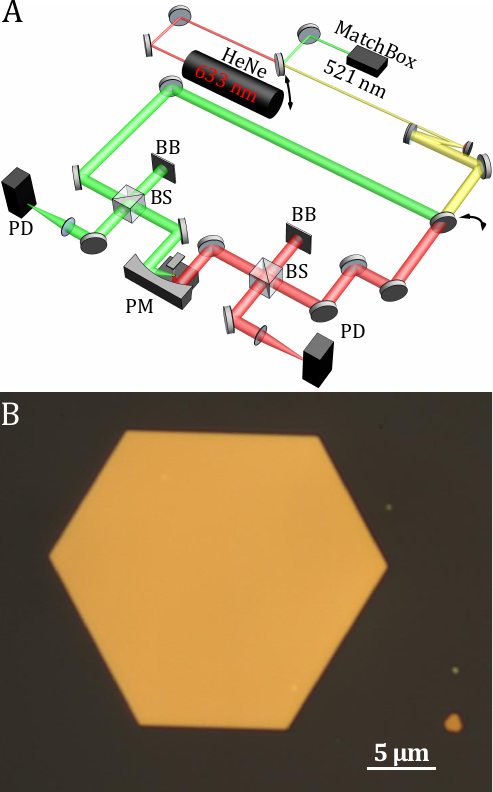}
    \caption{(A) Sketch of the near-field setup (PD$=$photo diode, PM$=$parabolic mirror, BS$=$beam splitter, BB$=$beam blocker), (B) optical microscope picture of platelet. \label{figure1}}
\end{figure}

The setup is based on a commercially available s-SNOM system (Neaspec Company, Munich, Germany). A sketch of the near-field setup is shown in Fig.~\ref{figure1}. The pilot laser of the system is a HeNe laser (Thorlabs) providing CW radiation at 633~nm. By using flip mirrors (curved arrows) we can couple in CW radiation from a 521~nm diode laser (INTEGRATED OPTICS). The respective beam is sent into a beam expander (4x) and can be coupled into the s-SNOM system via two beam paths. One beam path is entering the s-SNOM from the right-hand side. After the beam splitter the incident light is focused by a parabolic mirror (PM, NA=0.39) onto the tip with a polar angle of $\vartheta=60^\circ$ (see Fig.~\ref{figure3}). The back-scattered light is recollected by the PM and sent towards the Si photo receiver (New Focus). A second beam path for the 521~nm light enters the s-SNOM from the left-hand side (Fig.~\ref{figure1}A). Also here, the back-scattered light is sent towards the detector. The AFM probe tips are metal-coated commercial tips (Arrow NCPt, NanoWorld, Switzerland), made from single-crystalline Si and coated with a platinum-iridium (PtIr5) layer on both sides of the cantilever.

Figure~\ref{figure1}B shows an optical microscope picture of a platelet sample. The gold platelets have been produced by a wet-chemical synthesis process and were deposited on a glass substrate. The hexagonal structure features a high aspect-ratio with a lateral size of approximately 20 $\upmu$m and a height of 80-120~nm. The samples possess $\{111\}$ top and bottom facets and a composition of alternating $\{100\}$ and $\{111\}$-planes on the side facets. The typical RMS surface roughness is on the order of 200 pm. Therefore, the surface quality is very high with almost no defects or other impurities and the edges of the gold flakes are well defined, which, as we will show in the following, are essential for an undisturbed propagation of SPPs. More details about the synthesis process and the properties of the platelets can be found in a previous paper \cite{krauss2018controlled}.

\section*{SPP signal channels in SNOM}

As briefly introduced above, it is well known that SPPs on conductive surfaces can be excited by illumination of edges, (periodic) grooves, or other obstacles with light. Due to the strong field confinement at the apex of a metal-coated tip, also the illuminated AFM tip itself can serve as a source for the excitation of SPPs in s-SNOM. For metallic surfaces, this effect was first described by Chang \textit{et al.} \cite{chang2008fourier}, who studied SPPs generated by a single nanohole and nanohole arrays in silver using a illumination wavelength of $\lambda_0=$532 nm. In another study, Yan Li \textit{et al.} investigated SPPs on a thin gold layer excited by visible light ($\lambda_0=633 nm$) at a nanoslit \cite{li2014origin}. They showed that the obtained periodic field distribution can be explained by the coherent interference of the SPPs excited at the nanoslit and the incident light at the tip. However, in this study no contribution of tip-launched SPPs has been found and the signal vanished after 2.5 $\upmu$m propagation from the slit. Recently, Walla \textit{et al.} performed similar experiments by studying SPPs on a thin gold layer in the vicinity of a $10\times 10$~$\upmu$m$^2$ square hole using a illumination wavelength of $\lambda_0=800$ nm. With this sample geometry, they were able to observe SPP interference fringes in four different directions up to 5 $\upmu$m away from the edges of the square hole. They revealed that the excitation of SPPs at the tip in the visible is anisotropic and not circularly symmetric as it was considered in several studies before, and that this anisotropy is dependent on the geometry of the tip. Furthermore, they found an additional tip-mediated excitation channel for SPPs, where the incident light is first reflected at the tip before it hits an edge where SPPs are launched. We investigate the interference pattern obtained by s-SNOM at the surface of high quality gold platelets. Due to the hexagonal structure of the platelet it is possible to scrutinize the fringes from six different directions.

\begin{figure*}[htb]
\centering
	\includegraphics[width=0.78\textwidth]{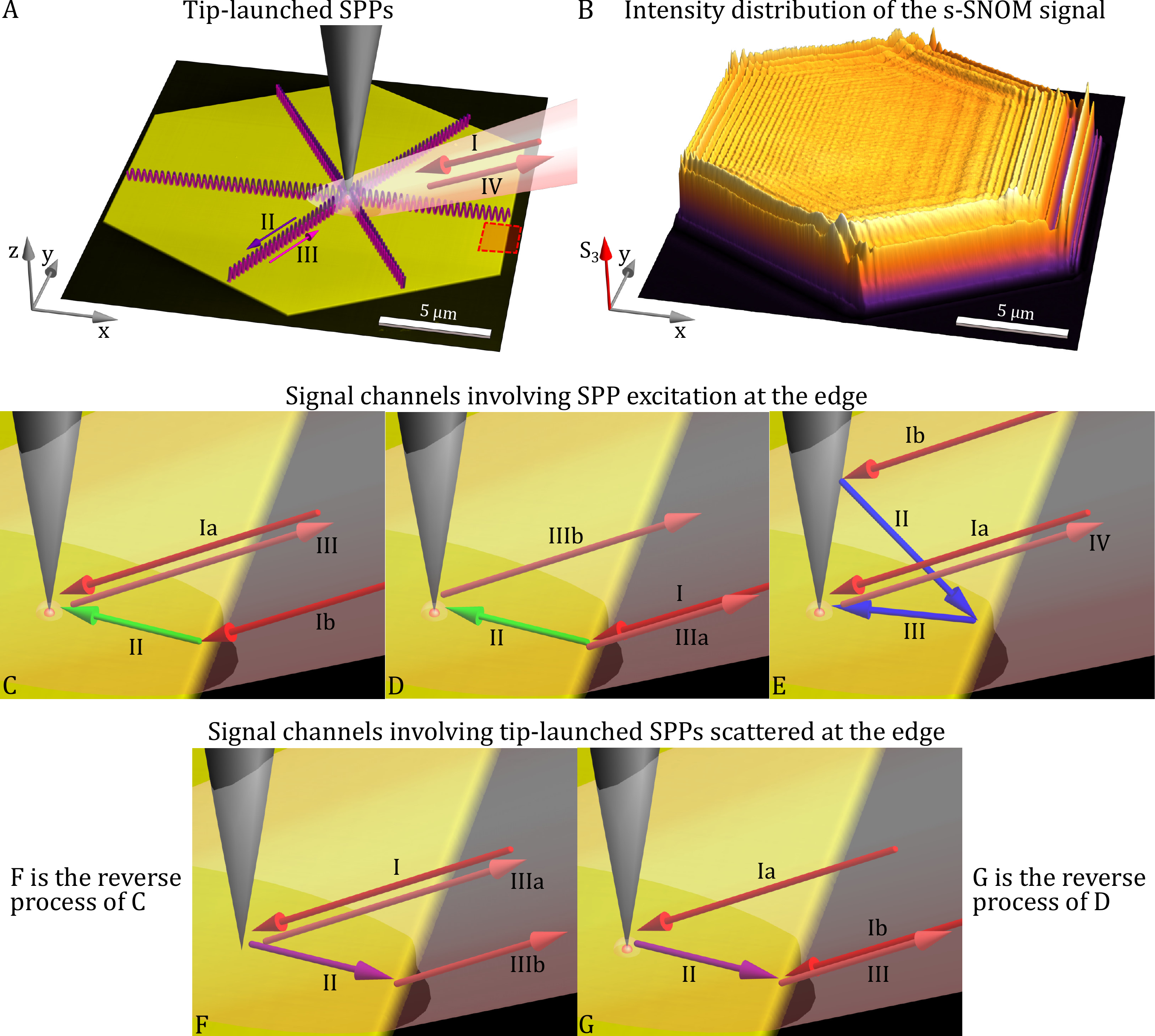}
    \caption{(A) AFM topography of an mono-crystalline gold platelet and illustration of tip-launched SPPs, (B) 3D-Plot of the intensity distribution of the s-SNOM signal, (C) illustration of edge-launched SPPs which interfere with the backscattered light from the tip, (D) illustration of edge-launched SPPs which interfere with the backscattered light from the edge, (E) illustration of tip-reflected edge-launched SPPs, (F\&G) illustration of the reverse signal channels shown in C and D, respectively, in which tip-launched SPPs are scattered into free space in the direction of the detector; these signal channels yield the same fringe spacing as C and D but with different intensities. The respective positions of the view windows shown in C-G are indicated by the red dashed box.} \label{figure2}
\end{figure*}

In the following, we summarize the different signal channels contributing to the interference patterns and connect them to models that have been used in the previous publications to estimate the corresponding fringe spacing. Non-interferometric s-SNOM raster scans (here using the HeNe laser at 633~nm) yield simultaneously the topography (Fig~\ref{figure2}A) and a map of the s-SNOM signal (Fig.~\ref{figure2}B). The intensity distribution of the scattered field shows that the signal is confined at the surface of the platelet. The complex interference pattern is formed by lines of maxima and minima with different spacings which are parallel to the edges. These fringes are the result of the interferences of different scattering pathways.

The formation of the interference pattern can be understood by a simple model. The detected signal at a certain tip position $\vec{x}$ is an intensity which is given by the sum of the individual scattered E-fields multiplied with the sum of their complex conjugated E-fields:

\begin{eqnarray}
I_{tot}(\vec{x})&=&(E_1(\vec{x})+E_2(\vec{x})+E_3(\vec{x})+...)(E_1^*(\vec{x})+E_2^*(\vec{x})+E_3^*(\vec{x})+...)\\
                &=&\sum_{i,j}{E_{i}(\vec{x})E^*_{j}(\vec{x})},
\end{eqnarray}

\noindent and it can be shown when expressing $E_{i/j}=\left|E_{i/j}\right|e^{i\phi_{i/j}}$ that

\begin{equation}
I_{tot}(\vec{x})=\sum_{i,j}\left|E_i(\vec{x})\right|\left|E_j(\vec{x})\right|\cos{\left(\Delta\phi_{i,j}(\vec{x})\right)},
\end{equation}

\noindent where $\Delta\phi_{i,j}=(\phi_j-\phi_i)$ is the phase difference between the E-field components $E_i$ and $E_j$ (see Supplemental A). In non-interferometric scans, the scattering pathway contributing the strongest to the detected signal is the incident light which is directly scattered back towards the detector $E_{bs}$. Furthermore, there are several SPP related scattering pathways which are due to the excitation and scattering of SPPs at the tip and/or an edge. Thus, we can order the contributions to the measured signal  $I_{tot}$ with respect to their respective strength the following way:

\begin{equation}
I_{tot}(\vec{x})=\left|E_{bs}\right|^2+2\sum_k{\left|E_{bs}\right|\left|E_k\right|(\vec{x})\cos{\left(\Delta\phi_{bs,k}(\vec{x})\right)}}+\sum_{l,m}{\left|E_l\right|(\vec{x})\left|E_m\right|(\vec{x})\cos{\left(\Delta\phi_{k,l}(\vec{x})\right)}},
\label{SigCont}
\end{equation}

\noindent where k, l and m are SPP related contributions. The measured intensity is given by a dominating constant offset $|E_{bs}|^2$, the next stronger contribution is given by the superposition of the homodyne amplified signals $E_k(\vec{x})$ , followed by the interference among different SPP related scattering pathways ($E_l(\vec{x})$ and $E_m(\vec{x})$). Note, that this also holds for the demodulated signal \cite{knoll2000enhanced}. The latter two contributions are modulated by the phase differences $\Delta\phi$ between the respective two fields and give rise to the fringes in the s-SNOM image of Fig.~\ref{figure2}B. All these contributions can be complex and difficult to disentangle, therefore, for clarity, we focus our analysis of the interference patterns on the stronger signal channels of the homodyne amplified scattering pathways. We will see in the following that this approach is justified by the fact that it gives rise to the most prominent contributions.

The first SPP related signal channel is given by tip-launched SPPs. As shown in Fig.~\ref{figure2}A, the incident beam (I) illuminates the tip, thereby launching SPPs under the tip apex. The SPPs propagate towards the boundaries of the hexagonal sample (II) and are reflected at the edges. In case the propagation direction is perpendicular to the respective edge, the reflected SPPs will propagate back towards the tip (III). When they reach the tip the SPPs are interfering with the tip-launched SPPs and scatter into free space (IV), where the scattered radiation can be measured. We call SPPs contributing to this signal channel \textit{tip-launched} (\textit{tl}) SPPs. For a certain distance $d$ between the tip and the edge the SPPs acquire a phase difference $\Delta\phi$ with respect to the exciting, incident light when propagating towards the edge and back to the tip. If the SPPs acquire a phase difference of $\Delta\phi=2\pi$, then the distance $d$ corresponds to the periodicity $\Lambda_{tl}$ of the fringes which is given by 

\begin{equation}
\Lambda_{tl}=\frac{\lambda_{SPP}}{2},
\end{equation}

\noindent
where $\lambda_{SPP}$ is the wavelength of the SPP. The tip-launched SPPs are damped, cylindrical waves and the amplitude behaves like $\left|E_{tl}\right|\propto \frac{1}{\sqrt{x}}e^{-\frac{\Gamma}{2}x}$ \cite{woessner2015highly}, where $\Gamma$ corresponds to the inverse of the propagation length $L$.

\begin{figure}[htb!]
\centering
	\includegraphics[width=0.5\textwidth]{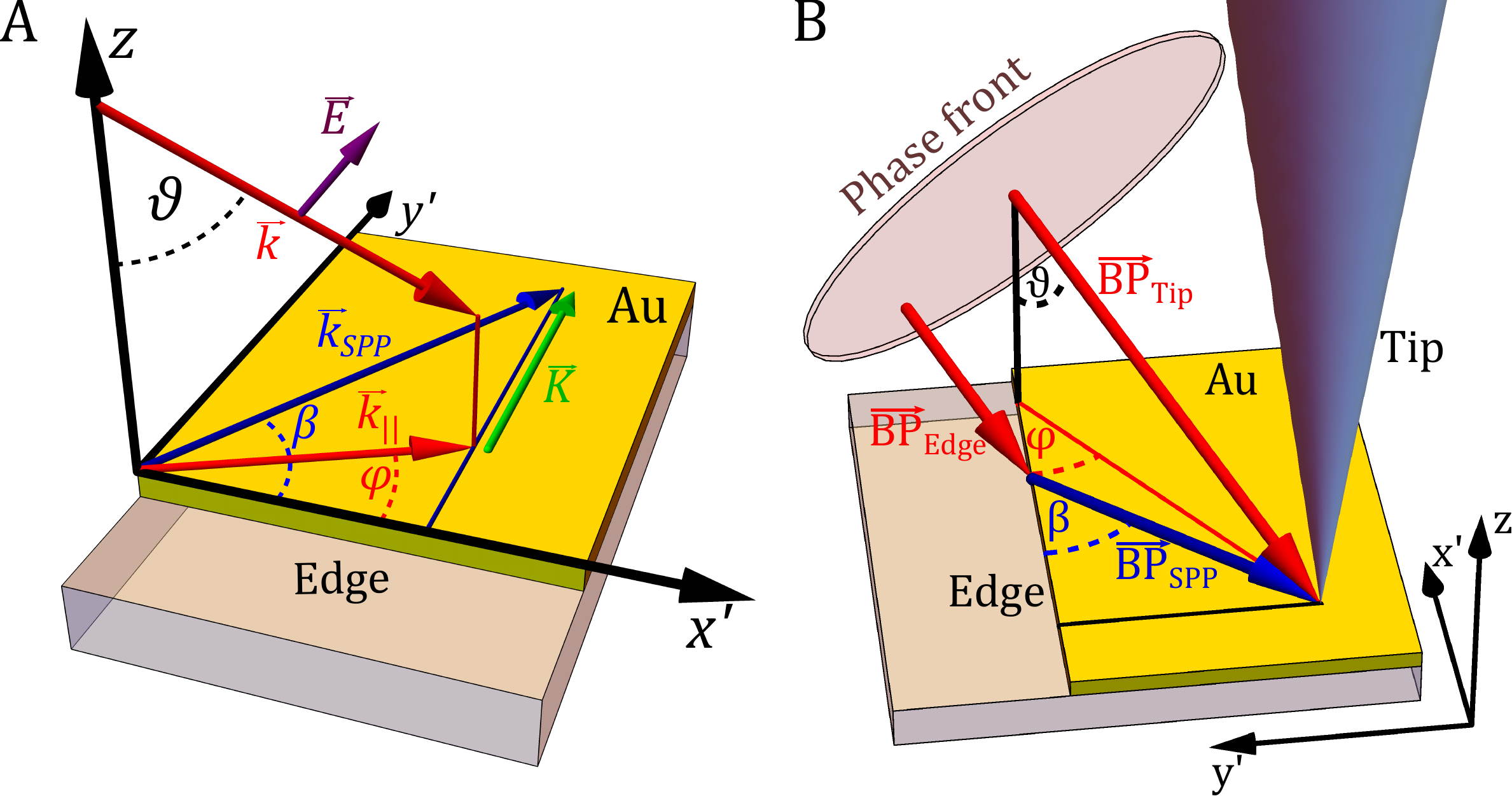}
    \caption{(A) Schematic of wave vectors and angles for the derivation of the fringe spacings of edge-launched SPPs, (B) schematics of beam path of incident light and edge-launched SPPs.\label{figure3}}
\end{figure}

Further signal channels are connected to edge effects, as can be seen by the strong modulation of the interference pattern close the edges in Fig.~\ref{figure1}B. As soon as the outer part of the focused incident beam hits one of the edges (Fig.~\ref{figure2}C), \textit{edge-launched} (\textit{el}) SPPs can be excited (Ib). Note that for the following discussion we are assuming a plane wave illuminating the tip and sample, although the maximal half-angle of the focused light cone is $\arcsin{(\textnormal{NA})}$$\approx$$23^\circ$, which is valid for the estimation of an average value. Furthermore, the spot size of the illuminating focused beam is limited and due to the shallow incident angle elongated along the incident axis. The propagation direction of the excited SPPs is depending on both the polar angle $\vartheta$ of the incident beam and the azimuth angle between the incident beam and the respective edge $\varphi$ as illustrated in Fig.~\ref{figure3}. The propagation direction is thereby determined by the phase matching condition at the edges, i.e. $k_{x'}=k_{SPP,x'}$, and the corresponding azimuthal angle $\beta$ of the SPP can be calculated from a generalized Snell's law:

\begin{align}
    & \cos{\left(\varphi\right)}\sin{\left(\vartheta\right)}=\mathfrak{Re}\left\{\tilde{n}\right\}\cdot \cos{\left(\beta\right)}\label{Eq:Snell1}\\
	\Leftrightarrow \; & \beta(\varphi,\vartheta)=\arccos{\left(\frac{\cos{\left(\varphi\right)}\sin{\left(\vartheta\right)}}{\mathfrak{Re}\left\{\tilde{n}\right\}}\right)}\label{Eq:Snell2}
\end{align}

\noindent where $\tilde{n}$ is the complex-valued effective refractive index of the SPPs (see also next section for more details). Parts of the edge-launched SPPs will eventually hit the tip (II), interfere with the incident light (Ia) and scatter into free space (III). The spacing of the resulting interference fringes can be obtained by rigorously calculating the acquired phase difference along the beam paths (BPs) of the incident light and the SPPs toward the tip illustrated in Fig.~\ref{figure3}B. As shown in this figure, one beam path points directly from the phase front towards the tip ($\overrightarrow{\textnormal{BP}}_{Tip}$) and the acquired phase is given by ${\phi_1=k_0 |\overrightarrow{\textnormal{BP}}_{Tip}|}$. The other beam path is divided into two parts. First, a ray of the incident light is propagating from the phase front towards the edge ($\overrightarrow{\textnormal{BP}}_{Edge}$). When it hits the edge with the lateral angle $\varphi$, SPPs are launched in a direction determined by $\beta$ and propagate towards the tip ($\overrightarrow{\textnormal{BP}}_{SPP}$). The acquired phase along this path is $\phi_2=k_0 |\overrightarrow{\textnormal{BP}}_{Edge}|+k_{SPP} |\overrightarrow{\textnormal{BP}}_{SPP}|$. If the distance $d$ between the tip and the edge of the platelet changes by a distance that equals to the fringe period ($\Lambda_{el1}$), the phase difference $\Delta\phi=\phi_2-\phi_1$ will change by $2\pi$. Thus, we obtain the following expression for the fringe spacing 

\begin{align}
     \Lambda_{el1}(\vartheta,\varphi)&=\frac{2\pi d}{\Delta\phi} \nonumber \\ 
																		&=\frac{\lambda_0}{-\sin{(\vartheta)}\sin{(\varphi)}+\sqrt{\sin^2{(\vartheta)}\sin^2{(\varphi)}-\sin^2{(\vartheta)}+n^2}},
\label{eq:2}
\end{align}

\noindent which is in accordance with the equation reported in Ref.~\cite{walla2018anisotropic}. Edge-launched SPPs can in good approximation be considered as damped, plane wave with the amplitude $\left|E_{el}\right|\propto e^{-\frac{\Gamma}{2}x}$.

Our measurements show that an additional edge-related effect which has yet not been addressed in the recent publications about SPPs on metallic surfaces by s-SNOM contributes to the interference pattern. Not only the tip, but also the edge of the monocrystalline gold platelet can scatter the incident light directly back to the detector and leads to a homodyne amplification of a SPP related scattering pathway. As an example, illustrated in Fig.~\ref{figure2}D, the incident beam (I) launches SPPs at the edge which propagate towards the tip (II), and scatter from the tip into free space (IIIb). The scattered SPP signal interferes with the part of the incident beam that is scattered back towards the detector from almost the same position at the edge as where the SPP is excited (IIIa). This effect only occurs when the edge is almost perpendicularly facing towards the incoming beam (here on the right-hand side), because only then the incident light hitting the edge will be partly scattered back within the collection angle of the PM and can interfere with the tip-scattered SPPs. From Figure~\ref{figure3}B one can see, that the acquired phase from the phase front to the edge and back is $\phi'_1=2 k_0 |\overrightarrow{\textnormal{BP}}_{Edge}|$. Along the beam path from the phase front to the edge, from the edge to the tip and from the tip to the phase front the phase accumulates to $\phi'_2=k_0 |\overrightarrow{\textnormal{BP}}_{Edge}|+k_{SPP} |\overrightarrow{\textnormal{BP}}_{SPP}|+k_0 |\overrightarrow{\textnormal{BP}}_{Tip}|$. Thus, the periodicity of the interference fringes can be calculated to:

\begin{align}
     \Lambda_{el2}(\vartheta,\varphi)&=\frac{2\pi d}{\phi'_2-\phi'_1} \nonumber \\ 
																		&=\frac{\lambda_0}{+\sin{(\vartheta)}\sin{(\varphi)}+\sqrt{\sin^2{(\vartheta)}\sin^2{(\varphi)}-\sin^2{(\vartheta)}+n^2}} \\
																		&=\Lambda_{el1}(\vartheta,-\varphi), \nonumber
\label{eq:4}
\end{align}

Another signal channel of SPPs is illustrated in Fig.~\ref{figure2}E and was first described in Ref.~\cite{walla2018anisotropic}. It also only contributes at edges which are facing towards the incident light. The incident beam is thereby partly impinging on the upper part of the tip (Ib) where the beam is eventually reflected towards the edge (II). Due to the reflection the path length (sum of the length of the blue arrows) is longer than in the case of edge-launched SPPs (green arrow in Fig.~\ref{figure2}C), which gives rise to an additional phase difference. Some part of the edge-launched SPPs will approach the tip (III), interfere with the incident beam (Ia) and scatter into free space (IV). These SPPs are called \textit{tip-reflected edge-launched} (\textit{trel}) SPPs. The periodicity of the observed fringes is in good approximation given by Eq.~\ref{eq:2}, when the polar angle $\vartheta$ of the incident beam is exchanged by the polar angle of the reflected beam $\vartheta'$ which has to be estimated from the measurement and is strongly dependent on the geometry of the tip used in the experiment:

\begin{equation}
\Lambda_{trel}(\vartheta',\varphi)=\Lambda_{el}(\vartheta',\varphi).
\label{eq:3}
\end{equation}

Finally, there are also reverse processes that can contribute to the s-SNOM image. Not only the tip, but also the edge can scatter SPPs to free space which have been observed, for instance, in a STM study \cite{zhang2013edge}. A similar effect was described by Hu \textit{et al.} investigating exciton-polariton transport in MoSe$_2$ waveguides by s-SNOM \cite{hu2017imaging}. In the first scenario, shown in Fig.~\ref{figure2}F, the incident light (I) launches SPPs at the tip that propagates towards the edge (II) where it scatters into free space (IIIb) and interfere with the directly back-scattered light from the tip (IIIa). This effect is described in Ref.~\cite{hu2017imaging}. In the second scenario, illustrated in Fig.~\ref{figure2}G, the incident beam (Ia) launches SPPs at the tip apex which propagate towards the edge (II), interfere with the part of the incident beam impinging on the same position at the edge (Ib) and scatter from the edge into free space (III). These effects contribute only at edges which are facing the incident beam, because only then the SPPs can scatter in the direction of the PM. By comparing the corresponding beam paths in Fig.~\ref{figure3}B it is evident that the acquired phase differences are equivalent to the two signal channels of edge-launched SPPs, respectively, and will produce the same fringe spacing as edge-launched SPPs. However, edge-launched and tip-launched SPPs are excited and scattered with different efficiencies and behave like plane waves and cylindrical waves, respectively. Therefore, the intensity of the measured signals will differ from each other. However, because they are overlapping it is difficult to disentangle these effects.

\section*{Experimental results and analysis}

\subsection*{Identification of interference processes}

In this s-SNOM study, we investigate the formation of the near-field intensity distribution of SPPs on platelets at two wavelengths in the red ($\lambda_{0,r}=633$~nm) and green ($\lambda_{0,g}=522$~nm) part of the visible spectrum. The respective wavelength of the SPPs is given by 

\begin{equation}
\lambda_{SPP}=\lambda_0/\mathfrak{Re}\left\{\tilde{n}\right\},
\label{eq:11}
\end{equation}

\noindent where $\lambda_0 $ is the free space wavelength of the incident light and $\tilde{n}=\tilde{k}/k_0=\sqrt{\tilde{\epsilon}/(1+\tilde{\epsilon})}$ is the complex-valued effective refractive index of the SPPs. We consider only SPPs propagating at the air-Au interface and neglect any hybridization effects with SPPs propagating at the SiO$_2$/Au interface. This is motivated by the sufficient thickness of our flakes (above 100 nm) \cite{olmon2012optical} and validated by our calculations (see Supplemental C). By using the data of Johnson and Christie \cite{johnson1972optical}, the complex-valued dielectric function of gold ($\tilde{\epsilon}$) at the two wavelengths yields the values $\tilde{\epsilon}(\lambda_{0,r})=-11.74+1.26i$ and $\tilde{\epsilon}(\lambda_{0,g})=-3.95+2.58i$. Note, that we are thus taking the non-neglectable effect of interband absorption into account, which is the limiting factor for both confinement and propagation length at our illumination wavelengths. We will use these analytically calculated values to compare to our experimental results. The respective theoretical values for the SPP wavelengths are thus $\lambda_{SPP,r}=606$~nm and $\lambda_{SPP,g}=477$~nm. The corresponding propagation lengths are given by $L=1/(2k_x'')$, where $k_x''=2\pi\mathfrak{Im}\left\{\tilde{n}\right\}/\lambda_0$, which yields $L_r=9.76$~$\upmu$m and $L_g=0.54$~$\upmu$m.

\begin{figure*}[ht!]
\centering
	\includegraphics[width=0.98\textwidth]{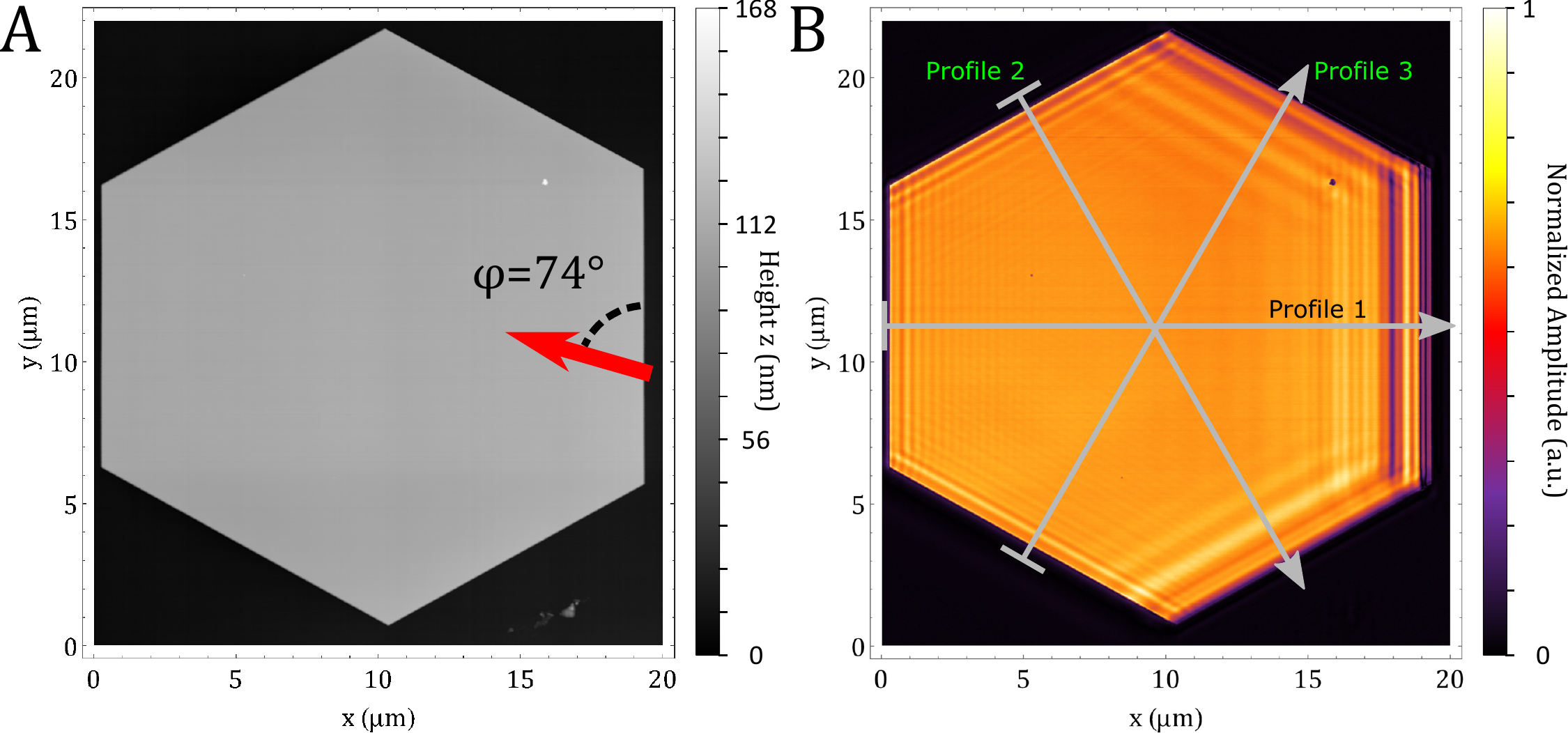}
    \caption{(A) AFM and (B) s-SNOM image of a platelet at an illumination wavelength of 633~nm.}
    \label{figure4}
\end{figure*}

Figure~\ref{figure4} shows (A) the AFM topography of a platelet and (B) the associated non-interferometric s-SNOM signal image taken at the 3$^{\textnormal{rd}}$ overtone of the tapping frequency of the s-SNOM ($\approx 285$ kHz) using the red HeNe laser at $\lambda_{0,r}$. The topography shows that the height of the platelet is around 110~nm and that the surface of the sample is very smooth (RMS surface roughness from AFM measurement $\approx 210$ pm) with almost no impurities. The red arrow indicates the angle of incidence of the laser beam. The sample has been oriented so that this angle is in the range of $75^\circ$ (here $\varphi=74^\circ$), where we got the clearest signal of SPPs due to the anisotropic excitation of SPPs at the tip apex \cite{walla2018anisotropic}. To minimize the scan area, it has been aligned with this edge by rotating the scan area by $29^\circ$ and the spatial resolution has been set to 25~nm. 

The image of the s-SNOM signal (Fig.~\ref{figure4}B) shows a pattern that is confined on the sample which proves that we are probing the near-field, also confirmed by the approach curves \cite{knoll2000enhanced}. It consists of lines of maxima and minima that are parallel to the edges, at which strong modulations are apparent. As we have described in the previous section, several signal channels overlap at the edges. Especially the excitation of edge-launched and tip-reflected edge-launched SPPs is strongly dependent on the position and orientation of the incident beam spot, which leads to the apparent anisotropic interference pattern at the edges and it is challenging to explain quantitatively the intensity pattern. Therefore, we focus on the investigation of the periodicity of these modulations. We extracted line profiles binned over 128 pixels along the three arrows with Gwyddion \cite{Necas2012}. Then, we performed a Fast Fourier Transform (FFT) to obtain the spatial frequency components ($K$) of theses profiles. However, we plot the spectra as a function of the fringe spacing $\Lambda=2\pi/K$, because this way the peaks for small fringe spacings become much more apparent.

\begin{figure*}[ht!]
\centering
	\includegraphics[width=0.98\textwidth]{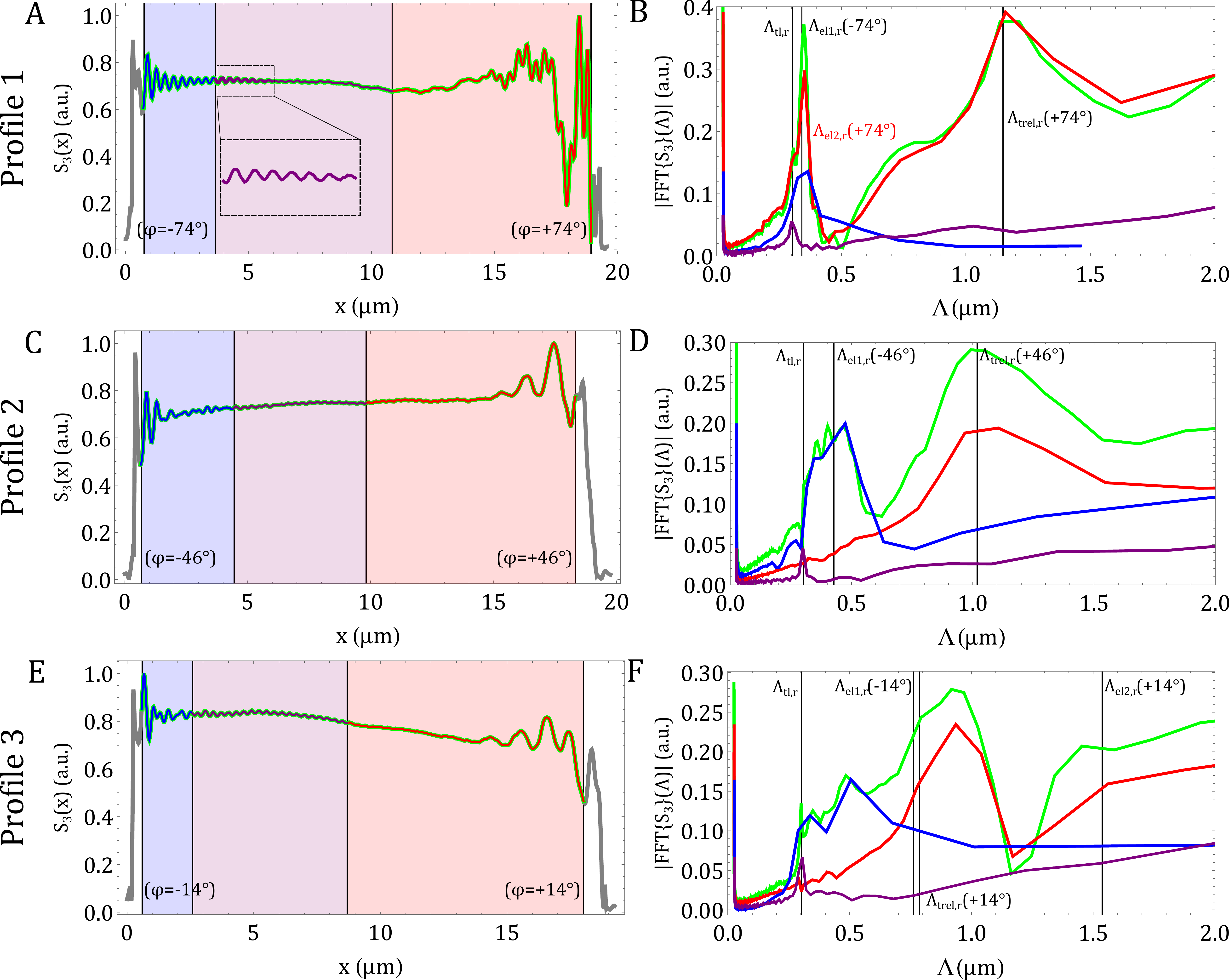}
    \caption{Profiles and corresponding FFTs from the s-SNOM scan at 633~nm shown in Fig.~\ref{figure4}B; (A) Profile 1, (B) corresponding FFT, (C) Profile 2, (D) corresponding FFT, (E) Profile 3, (F) corresponding FFT.}
    \label{figure5}
\end{figure*}

Figure~\ref{figure5}(A) shows the intensity profile 1 illustrated in Fig.~\ref{figure4}B. As defined in Fig.~\ref{figure3}, the angle subtended by the incident light and the corresponding edge $\varphi$ is $-74^{\circ}$ and $+74^{\circ}$. The profile rises steeply on the left-hand side where the platelet starts, then the signal oscillates and decays rather smoothly until the center of the flake. After approximately 10 $\upmu$m, a longer oscillation period appears which changes into a complex pattern of short fringe distances overlapped with a strong signal modulation. Finally, the signal decreases steeply at the other end of the sample. To analyze the spacing of the fringes we chose a range of the profile (green curve) and spans over the three colored regions in Fig.~\ref{figure5}A and performed a FFT over the entire range. The resulting spectrum plotted over $\Lambda=2\pi/K$ is shown in Fig.~\ref{figure5}B (green curve). The range has been set approximately 0.5-1 $\upmu$m away from the edges to avoid taking into account coupling effects between the tip and the edges \cite{Babicheva:17}. We calculated the expected values of the fringe spacings due to the effects described in the previous section and obtained $\Lambda_{tl,r}=0.303$~$\upmu$m, $\Lambda_{el1,r}(60^{\circ},-74^\circ)=0.347$~$\upmu$m and $\Lambda_{trel,r}(30^{\circ},+74^\circ)=1.149$~$\upmu$m where we found the best agreement when substituting $\vartheta=60^\circ$ with $\vartheta'=30^\circ$. The corresponding values are marked by the vertical lines in Fig.~\ref{figure5}B and we find that the maxima of the three peaks (green curve) overlap well with the calculated values. The predicted value for $\Lambda_{el1,r}(60^\circ,+74^\circ)=3.423$~$\upmu$m is masked by the DC-offset so we limit the plotting range to $\Lambda = 2.0$~$\upmu$m. To identify which part of the signal profile contributes to these peaks, we divided the profile in three parts marked by the light blue region (blue curve), the light purple region (purple curve) and light red region (red curve) in Fig.~\ref{figure5}A and performed a FFT of the s-SNOM signal on each interval. Note, that we performed the FFTs of the entire profile and the subintervals as they are shown without subtracting the background or using soft-windows. The corresponding spectra are also plotted in Fig.~\ref{figure5}B. The spectrum of the first interval (blue curve) close to the edge, where $\varphi$ is $-74^\circ$, peaks around the expected value for $\Lambda_{el1,r}(60^{\circ},-74^\circ)$. The broadness of the peak reflects the fact that edge-launched SPPs can only be excited as long as the incident beam hits the edge, furthermore tip-launched SPPs contribute to this signal. The spectrum of the second interval (purple curve) has one sharp peak centered almost exactly at the predicted value for tip-launched SPPs. We therefore conclude that this part of the profile is solely due to tip-launched SPPs. Impressively, this results show that we are able to measure a signal of the SPPs after they have propagated a distance of at least ${2\times 9\, \upmu\textnormal{m}=18 \, \upmu}$m because they are reflected at the edge, and we will use this signal, as we will show below, to determine the propagation length, which is defined as the distance for the SPP intensity to decay by a factor of $1/e$.
For the third subinterval we expected to obtain two peaks centered at $\Lambda_{tl,r}$ and $\Lambda_{trel,r}(30^\circ,+74^\circ)$. The broad peak at $\Lambda\approx 1.15$~$\upmu$m is in accordance with our estimation for tip-reflected edge-launched SPPs. We assign it to this signal channel because the oscillation with this period length is dominant between $x=10$~$\upmu$m and $x=15$~$\upmu$m. This range is far away from the edge and it is unlikely that the incident beam is still illuminating the edge at these tip positions. Thus, other edge-related signal channels cannot be considered. Furthermore, we find a dominant peak which has its maximum close to the predicted value for edge-launched SPPs which cannot contribute to this signal because they can only be excited on the other side of the sample ($\Lambda_{el1,r}(60^{\circ},-74^\circ)$). We therefore believe that this signal is due to edge-launched SPPs interfering with the incident light scattered at the same edge ($\Lambda_{el2,r}(60^{\circ},74^\circ)$). The slightly longer fringe spacing can be explained because the edge is not perpendicular with respect to the incident beam and the incident beam is not a plane wave but a light cone with the maximal half-angle of 23$^\circ$. Therefore, it is likely that a part of the incident light will be reflected/scattered by the edge within the collection angle of the PM and contribute to the interference signal with a slightly different phase difference. Another feature of this peak is a shoulder on its left side which is presumably due to tip-launched SPPs. 

The second profile marked in Fig.~\ref{figure4}B is shown in Fig.~\ref{figure5}C. For this profile the incident angle with respect to the two edges is $\varphi=\pm 46^\circ$. We applied the same procedure and use the same color coding for the curves and backgrounds as before. The predicted values for the angle-dependent fringe spacings are $\Lambda_{el1,r}(60^{\circ},-46^\circ)=0.428$~$\upmu$m and $\Lambda_{trel,r}(30^{\circ},+46^\circ)=1.018$~$\upmu$m for which we again substituted $\vartheta=60^\circ$ with $\vartheta'=30^\circ$. Also here the period for edge-launched SPPs from the right hand side $\lambda_{el1,r}(60^\circ,+46^\circ)=2.735$~$\upmu$m cannot be resolved by the FFT and thus is not shown. The spectrum of the entire range (green curve) consists of two broad peaks (Fig.~\ref{figure5}D). One peak is centred around the predicted value for edge-launched SPPs and shows indications for some contribution due to tip-launched SPPs on its left-hand side. The second peak is almost centred at the predicted value for tip-reflected edge-launched SPPs. The analysis of the subintervals reveals that the fringes close to the left edge, where $\varphi=-46^\circ$, can be attributed to edge-launched SPPs. The second subinterval has one clear peak centred at $\Lambda_{tl,r}$. Thus, also in this direction tip-launched SPPs can propagate over the entire platelet. The third subinterval can be related to tip-reflected edge-launched SPPs. In contrast to the third subinterval of profile 1, we do not obtain a strong peak close to $\Lambda_{el1,r}(60^\circ,-46^\circ)$, which is reasonable because it is unlikely that the incident light is reflected by the edge in the direction of the PM.

The results for the third profile in Fig.~\ref{figure4}B are presented in Fig.~\ref{figure5}E and F. The incident angle $\varphi$ is with $\pm14^\circ$ almost parallel to the respective edges. For this situation, the estimated values of the fringe spacings are $\Lambda_{el1,r}(60^{\circ},-14^\circ)=0.736$~$\upmu$m, $\Lambda_{el1,r}(60^\circ,+14^\circ)=1.592$~$\upmu$m and $\Lambda_{trel,r}(30^{\circ},+14^\circ)=0.797$~$\upmu$m. The spectrum of the entire range marked by the green curve (Fig.~\ref{figure5}F), peaks sharply at $\Lambda_{tl,r}$ and has a dominant and broad maximum at approximately 0.95~$\upmu$m, which is larger than the predicted values for both edge-launched SPPs from the left hand side and tip-reflected edge-launched SPPs from the right hand side. The amplitude rises again close to $\Lambda_{el1,r}(60^\circ,+14^\circ)$, but stays up because of the DC offset. The profile in the first selected subinterval (blue curve) shows a highly complex oscillation which quickly drops. The spectrum of this region starts to rise around the predicted value of tip-launched SPPs but does not show further features. The profile in the light purple window (purple curve) features a small but clear oscillation. Accordingly, the spectrum features a single peak at the estimated value for $\Lambda_{tl,r}$. Thus, also for this orientation we can clearly identify tip-launched SPPs which must have propagated for approximately 20~$\upmu$m. The oscillation in the third region (red curve) corresponds mainly to the peak around 0.95~$\upmu$m. It is actually surprising that also at this angle tip-reflected SPPs contribute to the signal because of the momentum conservation of the parallel component of the SPP wave vector, which means that the SPP should propagate away from the tip. However, it is likely that the incident beam will be scattered at the edge of the tip in a variety of directions, thus causing SPPs to be excited at the edges of the sample, that eventually propagate towards the tip and contribute to the signal. The width of the peak, as well as its dislocation are an indication that the process is not well-defined because of the steep angle of incident.   
The increase of the amplitude close to $\Lambda_{el1}(60^\circ,+14^\circ)$ is an indication for the contribution of edge-launched SPPs, but the resolution is too coarse for an exact assignment. Furthermore, we can see that in this region no signs for tip-launched SPPs can be identified. This is probably due to an inefficient generation of SPPs propagating in this direction.
In all spectra shown in Fig.~\ref{figure5} additional peaks and shoulders are apparent which do not overlap with the predicted values of the homodyne amplified signals (second term in Eq.~\ref{SigCont}). It is likely that these features are due to additional signal channels, especially due to the interference of different SPP related scattering pathways (last term in Eq.~\ref{SigCont}). However, these effect are relatively small compared to the homodyne amplified pathways, which justifies our approximation to neglect the higher order scattering pathways in this study. 

\begin{figure*}[!ht]
\centering
	\includegraphics[width=0.98\textwidth]{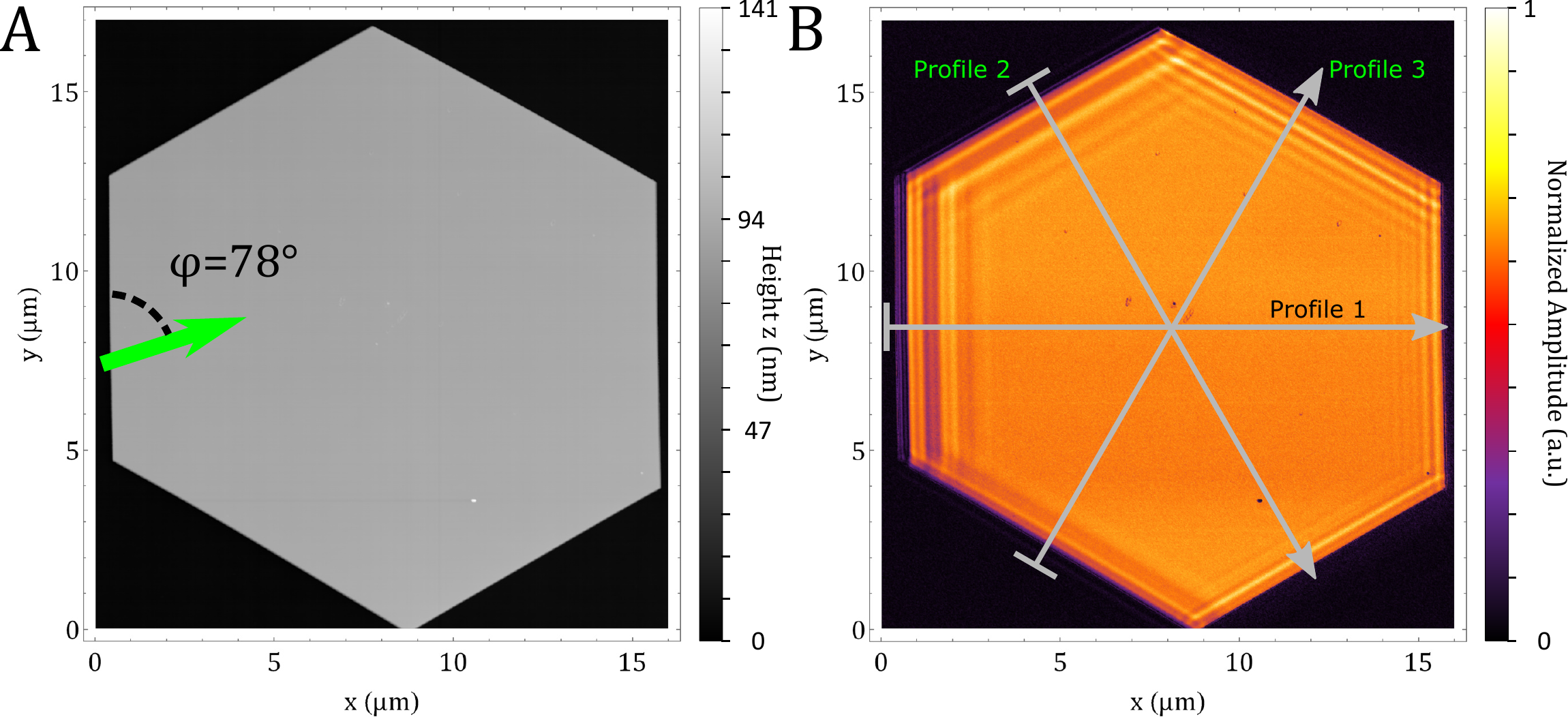}
    \caption{(A) AFM and (B) s-SNOM image of a platelet at an illumination wavelength of 521~nm.}
    \label{figure6}
\end{figure*}

We performed the same experiment with a different gold platelet with slightly smaller dimensions, using 521~nm excitation wavelength. The spacial resolution of this scan was set to 20~nm. The laser beam was coupled into the near-field setup from the left hand side, as indicated by the green arrow in Fig.~\ref{figure6}A. The surface of this sample is also of high quality and the height is constant at about 100~nm (RMS~surface~roughness~$\approx 230$~pm). The s-SNOM signal shown in Fig.~\ref{figure6}B is taken at the 4$^{\textnormal{th}}$ harmonic of the tapping frequency. Also here, the signal is confined on the sample and features interference fringes parallel to the edges whose spacing varies from edge to edge. In contrast to the result obtained at 633~nm, no interference fringes can be seen in the center of the sample which is due to the much shorter propagation length of SPPs excited by light at 521~nm ($L_g=540$~nm). We extracted three profiles along the arrows indicated in Fig.~\ref{figure6}B to analyze the contributions of the different SPP signal channels.

\begin{figure*}[ht]
\centering
	\includegraphics[width=1.0\textwidth]{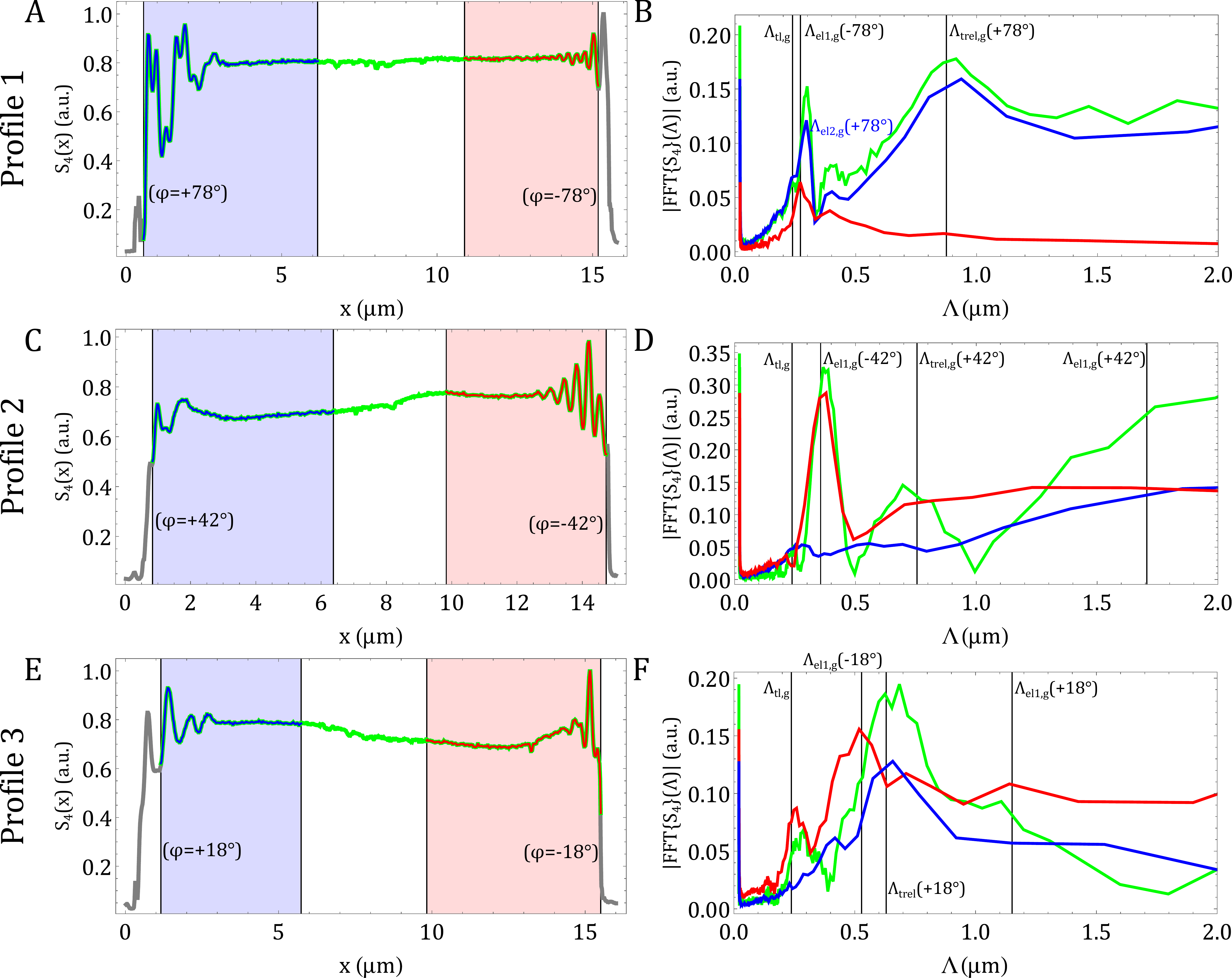}
    \caption{Profiles and corresponding FFTs from the s-SNOM scan at 521~nm shown in Fig.~\ref{figure6}B; (A) Profile 1, (B) corresponding FFT, (C) Profile 2, (D) corresponding FFT, (E) Profile 3, (F) corresponding FFT.}
    \label{figure7}
\end{figure*}

The first profile is presented in Fig.~\ref{figure7}A. The incident angle at the right- and left-hand side are $\varphi=\pm78^\circ$. We obtain the following predicted values for the fringe spacings: $\Lambda_{tl,g}=0.238$~$\upmu$m, $\Lambda_{el1,g}(60^\circ,-78^\circ)=0.270$~$\upmu$m, $\Lambda_{trel,g}(30^\circ,78^\circ)=0.869$~$\upmu$m for which we have also exchanged $\vartheta=60^\circ$ with $\vartheta'=30^\circ$~nm and $\Lambda_{el1,g}(60^\circ,+78^\circ)=2.243$~$\upmu$m which could not be resolved by the FFT and therefore is outside the plotted range. The most left and right vertical lines in Fig.~\ref{figure7}A mark the entire range we analyzed (green curve). The resulting FFT (Fig.~\ref{figure7}B, green curve) shows a small peak at $\Lambda_{tl,g}$, then it also peaks at a slightly larger fringe spacing than the predicted value for edge-launched SPPs at $\varphi=-78^\circ$ and features a broad peak at the estimated value for tip-reflected SPPs. Similar to profile 1 at 633~nm, the light blue subinterval which is at the edge facing the incident beam (blue curve, $\varphi=+78^\circ$) yields two peaks; one as expected for tip-reflected edge-launched SPPs $\Lambda_{trel,g}(30^\circ,+74^\circ)$ and a strong and narrower peak which is centered at a slightly larger fringe spacing than the predicted value for edge-launched SPPs excited at the opposite side of the sample where $\varphi=-78^\circ$. Therefore, we assign the latter peak to SPPs which are launched by the edge and propagate towards the tip, where they are scattered and interfere with the back-scattered incident light hitting the edge (see Fig.~\ref{figure3}E). In addition, this peak has a small plateau on the left-hand side which fits to tip-launched SPPs that are reflected at the edge. In the center of profile 1 we cannot find any signature of SPPs which is in accordance with the short propagation length. The FFT of the light red subinterval close to the edge where $\varphi=-78^\circ$ (red curve) has the expected peak centered at $\Lambda_{el1,g}(60^\circ,-78^\circ)$.

For the second profile shown in Fig.~\ref{figure7}C the incident angle at the two edges is $\varphi=\pm42^\circ$. The FFT of the entire range (green curve) has maxima at $\Lambda_{tl,g}$, $\Lambda_{el1,g}(60^\circ,-42^\circ)=256$~nm and $\Lambda_{trel,g}(30^\circ,+42^\circ)=0.750$~$\upmu$m, and the amplitude rises towards $\Lambda_{el1,g}(60^\circ,+42^\circ)=1.706$~$\upmu$m. The analysis of the signal at the left-hand side of the profile (light blue region) shows, however, no clear features that could be assigned to the expected fringe spacings. In contrast, the signal on the other side of the profile (light red region) can clearly be appointed to edge-launched SPPs where $\varphi=-42^\circ$ (Fig.~\ref{figure7}D).  

The results for the third profile, where $\varphi=\pm18^\circ$, are shown in Fig.~\ref{figure7}E and F. For these incident azimuth angles we estimate the following fringe spacings: $\Lambda_{el1,g}(60^\circ,-18^\circ)=0.528$~$\upmu$m, $\Lambda_{trel,g}(30^\circ,+18^\circ)=0.627$~$\upmu$m and $\Lambda_{el1,g}(60^\circ,+18^\circ)=1.151$~$\upmu$m. Also here, the FFT of the entire range (green curve) and the two subintervals at the respective edges yield peaks which can be assigned to the calculated fringe spacings of the respective signal channels.  As for the illumination at 633~nm, we find no clear contribution of tip-launched SPPs at the edge where $\varphi=+18^\circ$ (blue curve), however, the FFT yields a peak which can be assigned to tip-reflected SPPs, while the result for the other side of the profile ($\varphi=-18^\circ$) yields two peaks which fit to both tip-launched and edge-launched SPPs.
Also at this illumination wavelength we can find small peaks (especially in profile 1 and 3 around 0.4 $\upmu$m$^{-1}$) which are not overlapping with our predicted fringe spacings. These peaks are probably due to additional signal channels like interfering SPP related scattering pathways.

\subsection*{Determination of the SPP wavelength}

The results presented so far demonstrate that the Fourier analysis of the profiles allows to distinguish the contributing signal channels on a qualitative level. However, all effects which are related to the launching or scattering at the edges only contribute as long as the incident beam is also illuminating the edges. We believe that the finite beam spot size and the rather large half-cone angle of the focused beam are the reasons for the broadness of the edge-related peaks; the SPPs are excited at a range of positions, propagating in different directions than assumed in the case of an incident plane wave and still can contribute to the measured signal. This makes it difficult to retrieve further information about the propagation length and the actual wavelength of the SPPs.

\begin{figure}[htb]
\centering
	\includegraphics[width=0.75\textwidth]{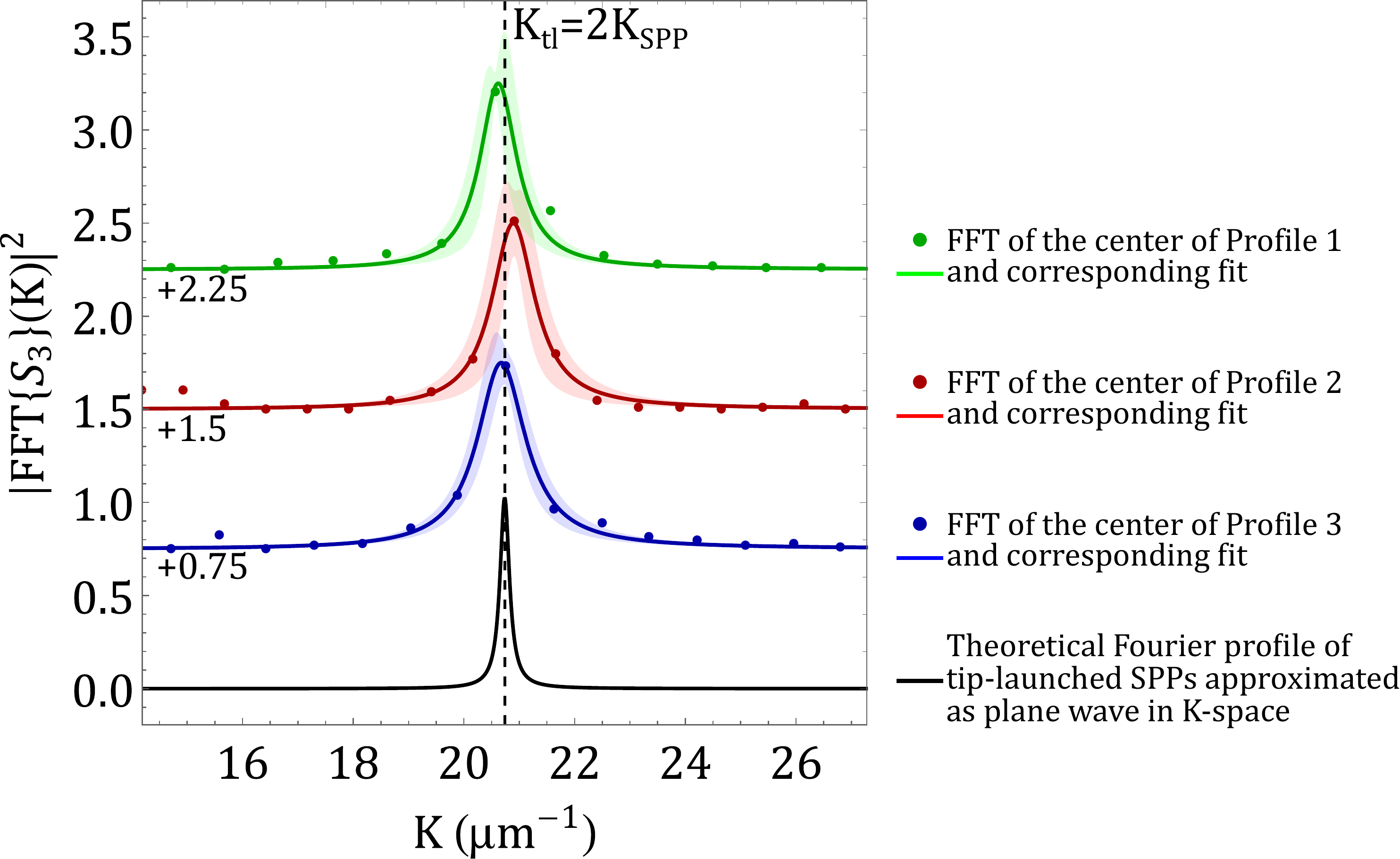}
    \caption{Squared FFT amplitude and corresponding fit of the 2$^\textnormal{nd}$ subinterval of the three profiles shown in Fig.~\ref{figure5} containing only the contribution of tip-launched SPPs (illumination wavelength at 633~nm).}
    \label{figure8}
\end{figure}

\noindent For the illumination at 633~nm, however, we can identify a range in the center of the platelets where the measured oscillations can unambiguously be related to tip-launched SPPs without any interfering contributions of other signal channels. These signatures can be used to extract the SPP wavelength $\lambda_{SPP}$. In order to do so, we take a closer look at the FFT of the second subintervals of the three profiles in Fig.~\ref{figure5}. Note, that we have chosen the interval to extend only to the center of the platelet to limit contributions of tip-launched SPPs which are reflected from the opposite side of the sample. As shown in Fig.~\ref{figure8}, where we plot the squared FFT amplitudes versus the spatial frequency $K$, all three profiles yield a clear rise of the signal amplitude close the estimated $K$-value of tip-launched SPPs. We mentioned before that tip-launched SPPs can be considered as cylindrical waves. With Eq.~\ref{SigCont} the detected intensity along a line perpendicular to the edge in the center of the platelet where we only find the contribution of tip-launched SPPs can be written as

\begin{equation}
I_{center}\approx \left|E_{bs}\right|^2+\frac{A}{\sqrt{2x}}e^{-\frac{\Gamma}{2}2x}cos{(K_{tl}x)},
\label{Eq:cylindrical}
\end{equation}

\noindent where the constant $A$ is taking the back-scattered E-field $\left|E_{bs}\right|$ and initial E-field of the launched SPP and reflection at the edge into account. The $2x$-terms are used because of the reflection of the SPP at the edge. Since we are already several micrometers away from the edge where the SPPs are reflected, we approximate Eq.~\ref{Eq:cylindrical} with a plane wave (see Supplemental B) which yields

\begin{equation}
I_{center}\approx \left|E_{bs}\right|^2+A e^{-\Gamma x}\cos{(K_{tl}x)}.
\label{Eq:PlaneWaveApprox}
\end{equation}

If we take now the Fourier transform of Eq.~\ref{Eq:PlaneWaveApprox}, the first, constant term results in a DC offset. The Fourier transform of the second term of Eq.~\ref{Eq:PlaneWaveApprox} can be solved analytically (in contrast to a cylindrical wave) and the absolute value squared of this Fourier transform can be written as

\begin{equation}
\left|\mathcal{F}\left\{A e^{-\Gamma x}\cos{(K_{tl}x)\Theta(x)}\right\}(K)\right|^2=\frac{A^2}{2\pi}\cdot \frac{\Gamma^2+K^2}{\Gamma^4+(K^2-K_{tl}^2)^2+2\Gamma^2(K^2+K_{tl}^2)},
\label{Eq:FFT}
\end{equation}

\noindent where $\Theta(x)$ is the Heaviside step function. In order to determine $K_{tl}$ we finally fit Eq.~\ref{Eq:FFT} to the absolute value squared of the FFT of the central profiles shown in Fig.~\ref{figure5}. The results of this procedure are presented in Fig.~\ref{figure8}. The colored shadings around the solid curves corresponds to the 95\% confidence level of the fit. The black curve shows the theoretical prediction given by Eq.~\ref{Eq:FFT}. Note, that the fit overestimates the decay rate $\Gamma$ of the SPPs because of the plane wave approximation so that no quantitative information can be deduced from that fitting parameter. However, from the fitting parameter $K_{tl}$, which corresponds to the peak position, we can calculate the SPP wavelength $\lambda_{SPP}=2\cdot 2\pi/K_{tl}$. We obtain the mean value of $\lambda_{SPP,exp}=606\pm 3$~nm for the SPP wavelength, which is in good agreement with the theoretical prediction. The fitting parameter of the decay rate yields a corresponding propagation length of $L_{exp}=2.1 \pm 0.3$~$\upmu$m, which is clearly too low, considering the fact that we can measure this signal channel 4-10 $\upmu$m away from the edge (this corresponds to 8-20 $\upmu$m of propagation). Note, that we used the errors on $K_{tl}$ and $\Gamma$ given by the 95\% confidence level to estimate the error of the mean values. Unfortunately, this analysis is not possible for the measurement taken with the green laser at 521~nm, because we cannot identify a range where tip-launched SPPs are isolated from edge-related effects. This emphasizes that it is important to consider edge-effects and other unintended excitations of SPPs when investigating plasmonic systems with s-SNOM.

\section*{Conclusion}

We have shown that the excellent surface quality of mono-crystalline gold platelets features superior plasmonic properties which allow accurate investigation of surface plasmons in the visible spectrum with s-SNOM in reflection mode. The fact that the plasmons are confined to the gold platelets, however, leads to the formation of complex s-SNOM images, which are particularly caused by edge effects. We have demonstrated a Fourier analysis approach to identify and separate the most prominent signatures of SPPs excited at the tip and the edges. Most importantly, we have been able to show that the wavelength of the SPPs can be determined if it is possible to isolate tip-launched SPPs. This was indeed possible at the illumination wavelength of 633~nm, where we have succeeded in estimating the SPP wavelength of 606 nm in good agreement with theoretical prediction.
We would like to emphasis that further work is necessary on the intensity and amplitude level, involving advanced FEM calculations to disentangle fully all the scattering pathways present at the edge of the platelet, as well as further development in the analytical methods to retrieve fully the propagation length. We believe that this work will trigger and pave the way toward such a realization. On the other hand, our study also shows that edge effects can be so dominant that it is difficult to study SPPs with a short propagation length such as at the excitation wavelength of 521 nm. Thus, it must be emphasized that the unintentional excitation of surface waves in s-SNOM measurements must always be considered in order to correctly interpret the results. Our reported methods and results will allow further investigations of high-quality plasmonic devices, paving the way towards future sensing, waveguiding and even quantum information processing applications in the visible range.

\section*{Acknowledgements}
The Center for Nanostructured Graphene is sponsored by the Danish National Research Foundation (Project No. DNRF103).

N.~A.~M. is a VILLUM Investigator supported by VILLUM FONDEN (Grant No. 16498).


\bibliographystyle{ieeetr}
\renewcommand{\refname}{REFERENCES}
\bibliography{refs}

\end{document}


\begin{singlespace}

\maketitle

\begin{flushleft}


$^{*}$\textbf{Corresponding Author}: Korbinian J. Kaltenecker, Nicoals Stenger, DTU Fotonik, email: \url{korkal@dtu.dk},  \url{niste@fotonik.dtu.dk}

\textbf{Keywords}: Plasmonics, near-field imaging, mono-crystalline gold

\today

\end{flushleft}

\end{singlespace}





\section*{Supplemental A - Analytical model of the origin of the intensity fringes}
Especially the interference of the directly back scattered light and the individual SPP related scattering pathway contribute to the observed fringes because of homodyne amplification which was already described by Knoll\&Keilmann in 2000 $\left[37\right]$. A simple model considering the different field contributions measured by the detector can be used to make this clear. The detected signal at a certain tip position $\vec{x}$ is an intensity which is given by the sum of the individual scattered E-fields multiplied with the sum of the complex conjugated E-fields:

\begin{equation}
I_{tot}=(E_1(\vec{x})+E_2(\vec{x})+E_3(\vec{x})+...)(E_1^*(\vec{x})+E_2^*(\vec{x})+E_3^*(\vec{x})+...)
\end{equation}

Note that the s-SNOM is primarily sensitive to the $z$-component of the E-field, i.e. the field component that is parallel to the tip axes. Expression can be written as:

\begin{eqnarray}
I_{tot}(\vec{x})&=&\sum_{i,j}{E_i(\vec{x})E_j^*(\vec{x})}\\
                &=&\frac{1}{2}\sum_{i,j}{E_i(\vec{x})E_j^*(\vec{x})}+\frac{1}{2}\sum_{i,j}{E_i(\vec{x})E_j^*(\vec{x})}\\
								                &=&\frac{1}{2}\sum_{i,j}{E_i(\vec{x})E_j^*(\vec{x})}+\frac{1}{2}\sum_{i,j}{E_i^*(\vec{x})E_j(\vec{x})}\\
                &=&\frac{1}{2}\sum_{i,j}{\left(E_i(\vec{x})E_j^*(\vec{x})+E_i(\vec{x})E_j^*(\vec{x})\right)}\\
                &=&\frac{1}{2}\sum_{i,j}{\left(\left|E_i(\vec{x})\right|e^{i\phi_i}\left|E_j^*(\vec{x})\right|e^{-i\phi_j}+\left|E_i(\vec{x})\right|e^{-i\phi_i}\left|E_j^*(\vec{x})\right|e^{i\phi_j}\right)}\\
                &=&\frac{1}{2}\sum_{i,j}{\left(\left|E_i(\vec{x})\right|\left|E_j^*(\vec{x})\right|e^{-i(\phi_j-\phi_i)}+\left|E_i(\vec{x})\right|\left|E_j^*(\vec{x})\right|e^{i(\phi_j-\phi_i)}\right)}\\
                &=&\frac{1}{2}\sum_{i,j}{\left|E_i(\vec{x})\right|\left|E_j^*(\vec{x})\right|(e^{-i\Delta\phi_{i,j}}+e^{i\Delta\phi_{i,j}})}\\
								&=&\frac{1}{2}\sum_{i,j}{\left|E_i(\vec{x})\right|\left|E_j^*(\vec{x})\right|\cos{\Delta\phi_{i,j}}}
\end{eqnarray}

\noindent where $\Delta\phi_{i,j}=(\phi_j-\phi_i)$ is the phase difference between the E-field components $E_i$ and $E_j$. The argument is now, that only these excitation and detection pathways are significantly contributing to the measured signal, which are interferences between the directly back-scattered light and the light from SPP related scattering pathways (\textit{i.e.}, tip-launched, edge-launched, tip-reflected edge-launched). The directly back-scattered light ($E_bs$) is stronger than the actual SPP signals. Thus, the (as we assume) constant back-scattered light $E_bs$ gives rise to a homodyne amplification of the interfering SPP signal $E_{SPP related}$. This argument holds also for the demodulated signals where a similar effect is even stronger as described in Ref.~$\left[37\right]$. For this reason, the contribution of the interferences between different SPP related pathways is weaker

\begin{equation}
\left(\left|E_{bs} \right|\left|E_{SPP related} \right|>\left|E_{SPP related} \right|\left|E_{SPP related} \right|\right).                                          
\end{equation}

Thus, we can order the contributions to the measured signal $I_{tot}$ with respect to their respective strength the following way:

\begin{equation}
I_{tot}(\vec{x})=\left|E_{bs}\right|^2+2\sum_k{\left|E_{bs}\right|\left|E_{jk}(\vec{x})\right|\cos(\Delta\phi_{bs,j}(\vec{x}))}+\sum_{l,m}{\left|E_l(\vec{x})\right|\left|E_m(\vec{x})\right|\cos(\Delta\phi_{l,m}(\vec{x}))},
\end{equation}

\noindent where $k$, $l$ and $m$ are SPP related contributions. As we can see, the measured intensity is given by a dominating constant offset $\left|E_{bs}\right|^2$, the next stronger contribution is given by the superposition of the homodyne amplified signals $E_k$, followed by the interference among different SPP related scattering pathways ($E_l$ and $E_m$). These individual contributions are modulated by the phase differences $\cos(\Delta\phi_{bs,j})$ and $\cos(\Delta\phi_{l,m})$ between the respective two fields which are dependent on the location of the excitation and scattering of SPPs at the tip and/or the edges. In this work, we focus on the modulation of the signal due to the homodyne amplification, because these contributions manifest in the most prominent peaks in the FFTs. For this reason, we think it is justified to focus our analysis on edge- and tip-related SPP scattering pathways:

\begin{equation}
I_{tot}(\vec{x})\approx\left|E_{bs}\right|^2+2\sum_{j}{\left|E_{bs}\right|\left|E_j(\vec{x})\right|\cos(\Delta\phi_{bs,j}(\vec{x}))}.
\label{eq:5}
\end{equation}

\section*{Supplemental B - Determination of SPP wavelength}

In the center of the flake we observe only one peak which can be related to tip-launched SPPs. Thus Eq.~\ref{eq:5} can be written in one dimension as:

\begin{equation}
I_{center}(x)\approx \left|E_{bs}\right|^2+2\left|E_{bs}\right|\left|E_{tl}(x)\right|\cos\left(\Delta\phi_{bs,tl}(x)\right).
\label{eq:6}
\end{equation}

Tip-launched SPPs are not plane waves, but cylindrical waves because they are launched from the tip which can be seen as a point source. Generally, the amplitude of tip-launched SPPs can be written as

\begin{equation}
\left|E_{tl}(x')\right|=\frac{A}{\sqrt{x'}}e^{-\frac{\Gamma}{2}x'}. 
\label{eq:7}  
\end{equation}

Note, that $\Gamma$ is defined as the inverse of the propagation length $L$. Since we are measuring the SPPs reflected at the edge and coming back to the tip (tl) as a function of the distance to the edge $x$, we thus have $x'=2x$, Eq.~\ref{eq:6} can be written with Eq.~\ref{eq:7} as

\begin{equation}
I_{center}(x)\approx \left|E_{bs}\right|^2+2\left|E_{bs}\right|\frac{A}{\sqrt{x'}}e^{-\frac{\Gamma}{2}x'}\cos\left(\Delta\phi_{bs,tl}(x)\right).
\label{eq:8}
\end{equation}

The Fourier transformation of Eq.~\ref{eq:8} cannot be written in a closed form and since we are not measuring directly from the edge but several micrometers away from the edge it cannot be used to extract $\Gamma$ exactly, because we are missing the information about the decay of the tip-launched SPP close to the edge.

\begin{figure*}[htb]
\centering
	\includegraphics[width=0.75\textwidth]{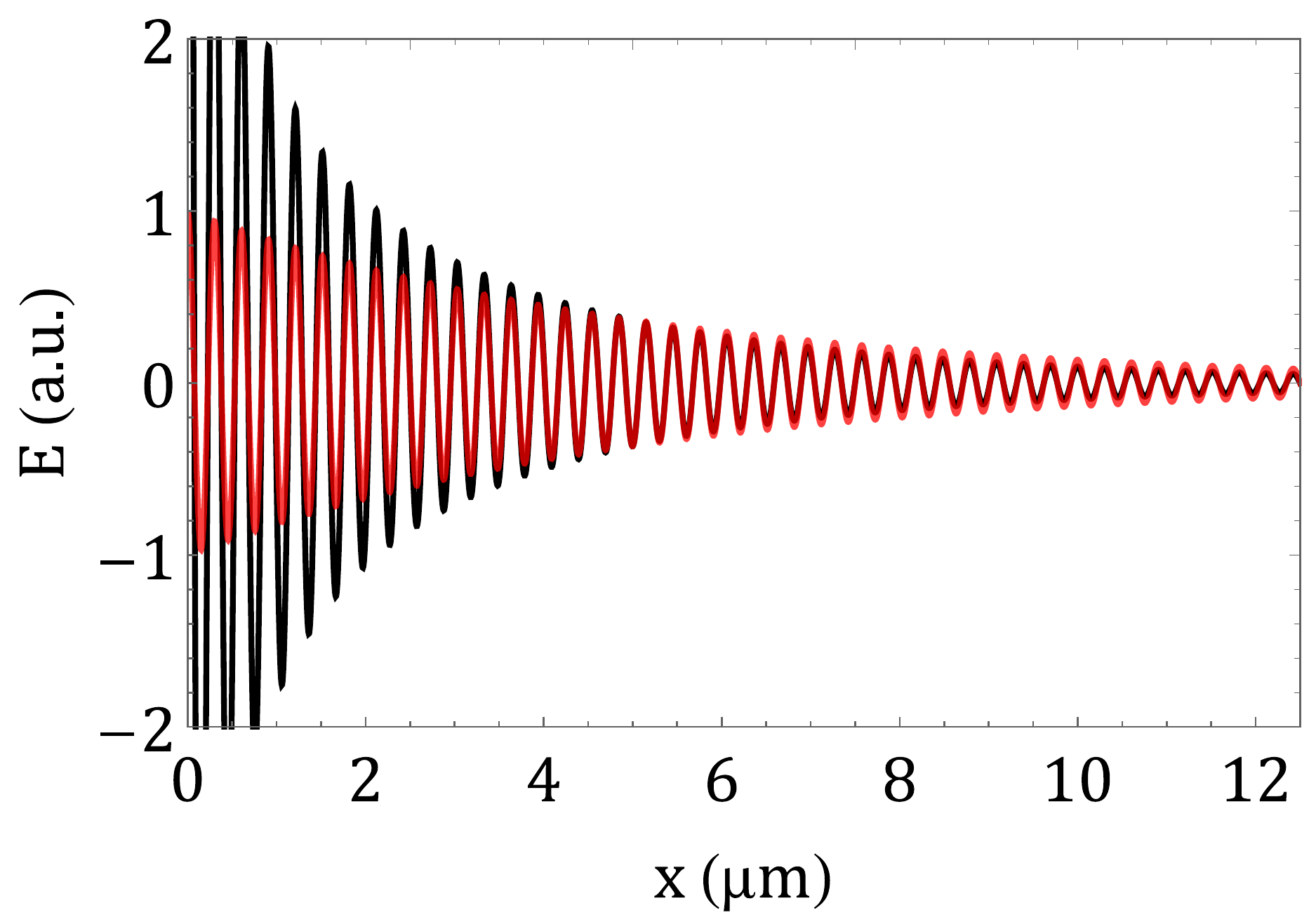}
    \caption{Comparison of cylindrical (black curve) and plane wave (red curve).} \label{figure1}
\end{figure*}

For this reason, we approximate the cylindrical wave in Eq.~\ref{eq:8} as a plane wave. In Fig.~\ref{figure1} we compare a cylindrical wave $(\frac{A}{\sqrt{2x}} e^{-\Gamma x}\cos(K_{tl} x))$ and a plane wave $(e^{-\Gamma x}\cos(K_{tl}x))$, where the amplitude $A$ of the cylindrical wave has been adjusted so that it overlaps at $x\approx 5$ $\upmu$m. While there is a huge difference for small $x$, the difference between 4 $\upmu$m and 12 $\upmu$m is small. However, by using the plane wave approximation we will clearly overestimate $\Gamma$, thus underestimate the propagation length of the SPP $L$, which means that we cannot deduce quantitative information from $\Gamma$ regarding the propagation length $L$.
We use now the following procedure: We have adjusted the interval in the center so that in each case these intervals start at $\approx 4$ $\upmu$m and end in the center. The reason is that at least in profile 1 and 3 (in Figure 5 in the main text) tip-launched SPPs are also coming from the other side of the sample and this way we can minimize their contribution. Then we perform the FFT of the intervals and take the squared absolute value. The absolute value squared of the Fourier transform of a plane wave yields

\begin{equation}
\left|\mathcal{F}\left\{e^{-\Gamma x}\cos{(K_{tl}x)\Theta(x)}\right\}(K)\right|^2\propto\frac{1}{2\pi}\cdot \frac{\Gamma^2+K^2}{\Gamma^4+(K^2-K_{tl}^2)^2+2\Gamma^2(K^2+K_{tl}^2)},
\label{eq:9}
\end{equation}

\noindent where $\Theta(x)$ is the Heaviside step function, and we fit Eq.~\ref{eq:9} to the FFT of our data. This way, we obtain $K_{tl}$ and $\Gamma$ and can calculate from there the fringe spacing $\Lambda_{tl}=\frac{2\pi}{K_{tl}}$ and from there the SPP wavelength $\lambda_{SPP}=2\Lambda_{tl}$.

\section*{Supplental C - Effect of the thickness of the gold platelets}

We have performed numerical calculations to determine the effect of the thickness of the gold platelet and the substrate and found that for a thickness of 75 nm and more the single interface expression is valid. As can be seen in Fig. R3, the difference at 75 nm at the illumination wavelengths we are using are marginal. Figure S2 show the result of the numerical simulations for a thickness of (a) 60 nm and (b) 75 nm. At 633 nm (1.96 eV) the difference between the dispersion relations of SPPs on a single interface of gold and air/silicon dioxide compared to a thin film are already neglectable for a thickness of 60 nm. At 521 (2.38 eV) nm this is the case form a thickness of 75 nm.

\begin{figure*}[htb]
\centering
	\includegraphics[width=1\textwidth]{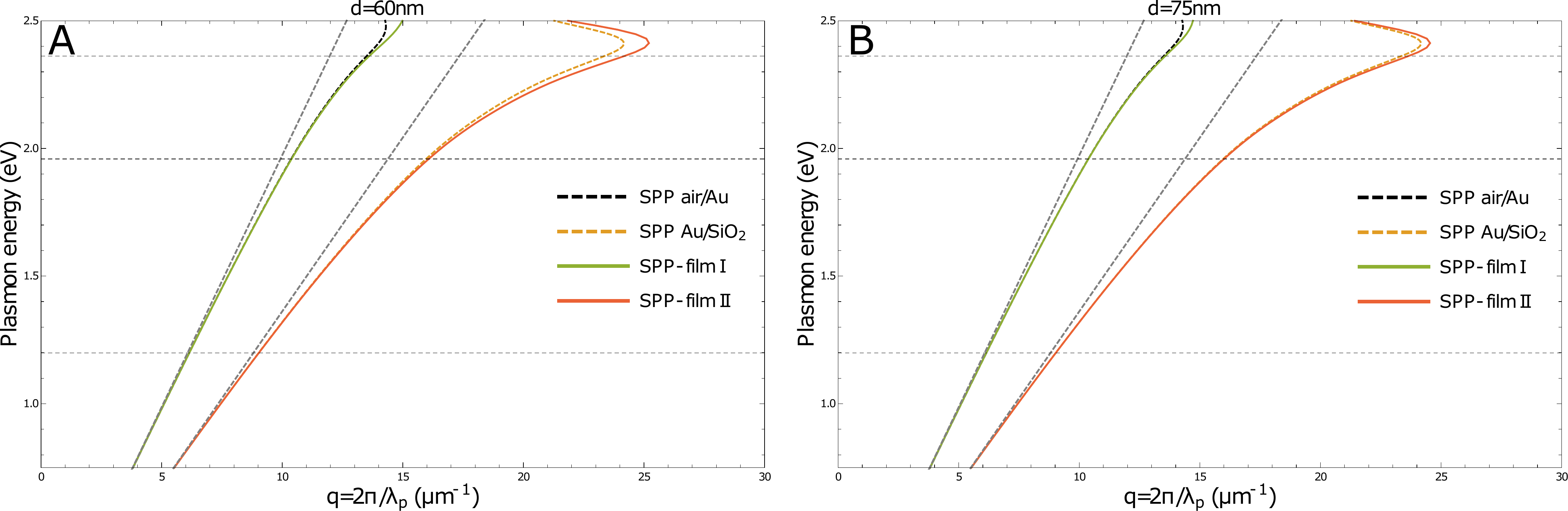}
    \caption{Comparison of the dispersion relation of SPPs on a single interface of gold and air/silicon dioxide and a thin film of (a) 60 nm and (b) 75 nm thickness.} \label{figure2}
\end{figure*}

\section*{Supplental D - SPPs propagation on monocrystalline gold platelet with rough surface}

\begin{figure*}[htb]
\centering
	\includegraphics[width=0.75\textwidth]{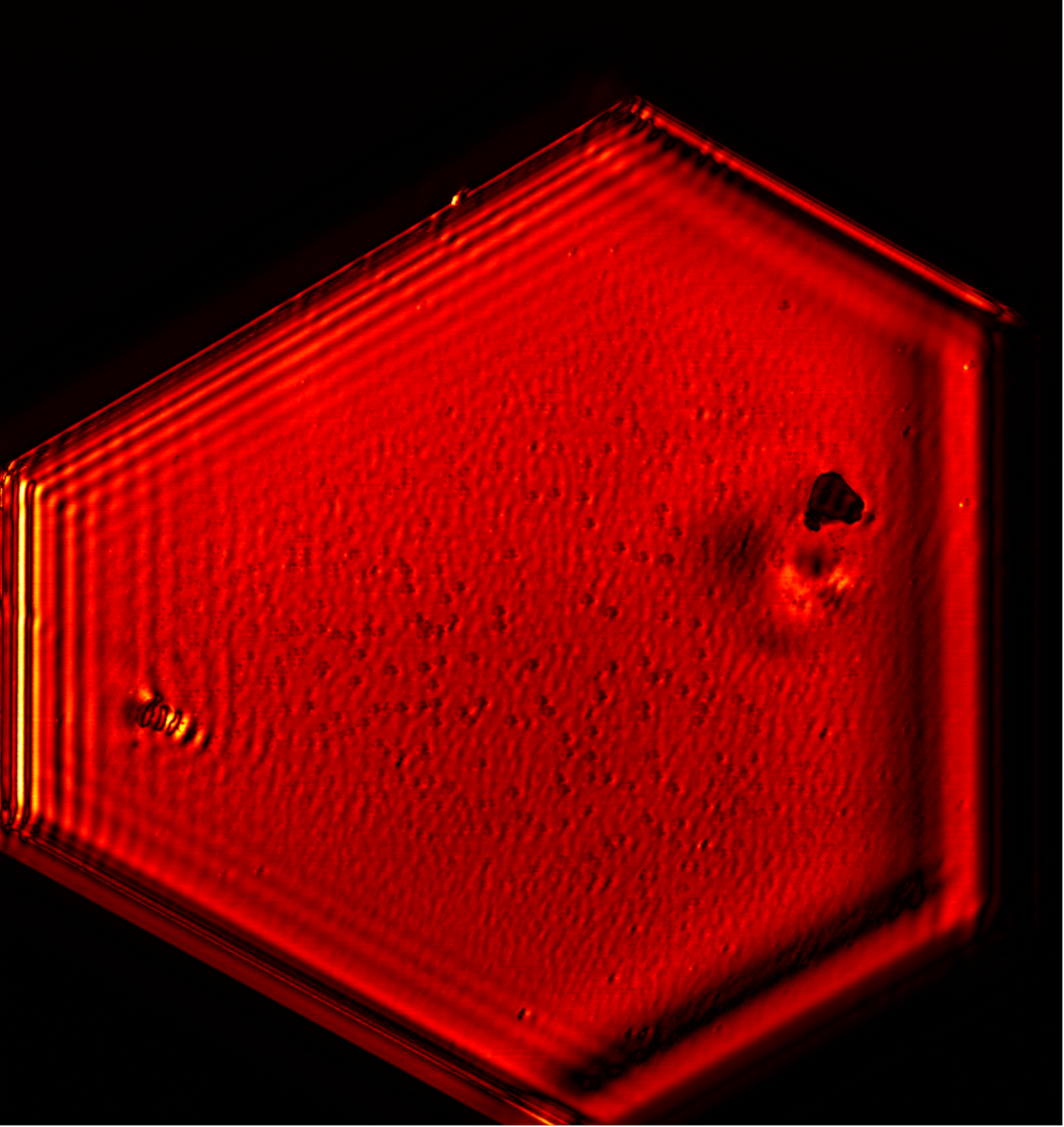}
    \caption{Plasma etched gold platelet, scan area 17 $\upmu$m $\times$ 18 $\upmu$m, the surface in the center of the platelet is rougher compared to the areas close to the edges. We cannot measure the typical SPP related fringes in the center because the SPPs are scattered.} \label{figure3}
\end{figure*}

We have carefully checked our claim that high quality metal surfaces are paramount in the study of SPP propagation by using Au flakes where the surface in the center has been significantly modified, displaying larger roughness, and measured them with our method. As it can be seen in Fig.~\ref{figure3}, the rough surface scatters the SPP and destroys the plane wavefronts of the SPP related signals. This makes the retrieval of the SPP properties in the center impossible.

